\documentclass[prd,twocolumn,aps,dvips]{revtex4}
\usepackage[english]{babel}
\usepackage{amscd}
\usepackage{epsfig}
\usepackage{tabularx}
\usepackage{verbatim}
\usepackage{graphicx}
\usepackage{latexsym}
\usepackage{amsmath}
\usepackage{amsfonts}
\usepackage{amssymb}
\usepackage[latin1]{inputenc}
\usepackage{color}

\newcommand{\be}{\begin{equation}}
\newcommand{\ee}{ \end{equation}}
\newcommand{\ben}{\begin{eqnarray}}
\newcommand{\een}{\end{eqnarray}}

\begin{document}

\title{Thermal properties of a solid through $q$-deformed algebra}

\author{Andre A. Marinho$^{a}$, Francisco A. Brito$^{b}$, Carlos Chesman$^{a,c}$}

\affiliation{$^{a}$ Departamento de Física Teórica e Experimental, Universidade Federal do Rio 
Grande do Norte, 59078-970 Natal, RN, Brazil\\
$^{b}$ Departamento de Física, Universidade Federal de 
Campina Grande, 58109-970 Campina Grande, Para\'\i ba, Brazil
\\
$^{c}$ Materials Science Division, Argonne National Laboratory, Argonne,
Illinois, 60439, USA
}



\begin{abstract} 

We study the thermodynamics of a crystalline solid by applying $q$-deformed algebras. We based part of our study on both 
Einstein and Debye models, exploring primarily $q$-deformed thermal and electric conductivities as a function of Debye 
specific heat. The results revealed that $q$-deformation acts as a factor of disorder or impurity, modifying the characteristics 
of a crystalline structure, as for example, in the case of semiconductors. 

\end{abstract}


\maketitle


\section{Introduction}
\label{int}

Quantum groups and quantum algebras can be considered $q$-deformations of Lie groups and semi-simple Lie algebras \cite{anat}. 
Physical systems with $q$-deformation have been the subject of intense research due to the emergence of quantum group 
structures caused by certain physical problems. This establishes a connection with $q$-analysis, known in mathematics 
for over a century. From the mathematical point of view, $q$-deformed algebra usually requires Hopf algebra. The $q$-deformation 
of a physical system is not the only way to deform it; rather, there are several ways of applying deformation in different physical 
and mathematical contexts \cite{sin,tsallis}.

For example, the widely studied harmonic oscillator system has several $q$-deformed descriptions \cite{bie, mac}. 
These are obtained from each other by transformation, as shown in \cite{borz, flo}. Some of the main problems in obtaining 
the $q$-oscillator are the spectrum, Hamiltonian, position and time operator. Moreover, the $q$-deformation of physical 
systems beyond the oscillator is not well studied. Most concepts of classical mechanics and quantum mechanics become 
unclear after being $q$-deformed \cite{day}.

This study extends our previous analysis \cite{bri} to solids. A solid consists of a large number of atoms linked by 
cohesive forces of various kinds. On the other hand, in a gas, molecules are free to move around the container, while in a 
liquid, they have less freedom, but still travel considerable distances. Atomic motion in a solid is very slight, causing 
every atom to move only within a small neighborhood, and vibrate around its equilibrium point. In a crystalline solid, 
the equilibrium points of atomic vibrations form a regular spatial structure, such as a cubic structure. 

Interaction between atoms allows the propagation of elastic waves in solid media, which can be both horizontal and longitudinal. 
If atomic oscillations around equilibrium positions are small, which should occur at low temperatures, the potential interaction 
energy can be approximated by a quadratic form of atomic displacements. A crystalline solid, whose atoms interact according to 
this potential, is called a harmonic solid. In harmonic solids, elastic waves are harmonics and the normal modes of vibration in 
crystalline solids \cite{kit}. A large number of phenomena involve quantum mechanical motion, in particular thermally-activated 
particles, obeying the $T^{3}$ law. Thermal excitations in the system are responsible for phonon excitation \cite{patt}. 

The study conducted by Anderson, Lee and Elliot \cite{and,lee,ell} shows that the presence of defects or impurities in a crystal changes 
the electrostatic potential in their neighborhoods, breaking the translational symmetry of the periodic potential. This disturbance 
can produce electronic wave functions located near the impurity, ceasing to be propagated  throughout the crystal. 

The conductivity of semiconductors can also be dramatically altered by the presence of impurities, i.e., different 
from atoms that make up the pure crystal. This property enables the manufacture of a variety of electronic devices from 
the same semiconductor material. The process of placing impurities of known elements in a semiconductor is called doping. 
Next we see the application of $q$-deformation in solids acting as a defect or impurity, specifically in Einstein and Debye 
solids, and observe the consequences; for example, temperature and thermal conductivity in the Debye model and other features of 
chemical elements \cite{kit}. We found interesting results, leading us to apply the same approach to other thermodynamic properties 
of solids, since this initial analysis shows that $q$-deformed elements have thermal properties are similar to another element in 
its pure state. 

The paper is organized as follows: in Sec. \ref{qdqa} we present the $q$-deformed algebra; in Sec. \ref{iqd} we implement 
$q$-deformation to Einstein and Debye solids in order to explore thermal properties; and finally, in Sec. \ref{con} we make our final 
comments. 
\section{$q$-Deformed quantum algebra}
\label{qdqa}
The $q$-deformed algebra of the quantum oscillator is defined by $q$-deformed Heisenberg algebra in terms of creation and 
annihilation operators $a^{\dagger}$ and $a$, respectively, and quantum number $N$ by \cite{chai,ng,sak},
\begin{eqnarray} [a,a] = [a^{\dagger},a^{\dagger}] =0\;,\qquad\qquad\ aa^{\dagger} - 
q^{-\frac{1}{2}}a^{\dagger}a= q^{\frac{N}{2}},\end{eqnarray} 
\begin{equation}[N,a^{\dagger}]= a^{\dagger}\;, \qquad\qquad [N,a] = -a ,\end {equation} 
where deformation parameter $q$ is real and the observed value of $q$ has to satisfy the 
non-additivity property  (see \cite{tsallis} for a comprehensive study on this property in many physical issues)
\begin{equation} [x+y] \neq [x] + [y].\end{equation}
In addition, the operators obey the relations 
\begin{eqnarray}[N]=a^{\dagger}a\;,\quad\qquad\qquad aa^{\dagger}=[1+N]. \end {eqnarray}
The $q$-Fock space spanned by orthornormalized eigenstates ${|n\rangle}$ is constructed according 
to 
\begin{eqnarray} {|n\rangle} =\frac{(a^{\dagger})^{n}} {\sqrt{[n]!}}{|0\rangle},\qquad\qquad 
a{|0>}=0 ,\end {eqnarray} 
The actions of $a$, $a^{\dagger}$ and $N$ on the states ${|n\rangle}$ in the $q$-Fock space 
are known to be 
\begin{equation} a^{\dagger}{|n\rangle} = [n+1]^{1/2} {|n+1\rangle},\end{equation} 
\begin{equation} a{|n\rangle} = [n]^{1/2} {|n-1\rangle},\end{equation} 
\begin{equation} N{|n\rangle} = n{|n\rangle}.\end{equation} 
The oscillator \cite{bie, bie1, mac} allows us to write the $q$-deformed Hamiltonian \cite{anat} as follows: 
\begin{equation} \label{eq1}{\cal H} = \frac{1}{2}\Big{\{a,a^{\dagger}\Big\}}.\end{equation} 
We have the \textit{basic $q$-deformed quantum number} $n$ defined as \cite{erns}, 
\begin{equation}\label{eq2} [x]=\frac {q^{\frac{x}{2}}-q^{-\frac{x}{2}}} 
{q^{\frac{1}{2}}-q^{-\frac{1}{2}}} \equiv[N]=a^{\dagger}a=\frac {q^{\frac{N}{2}}-q^{-\frac{N}{2}}} 
{q^{\frac{1}{2}}-q^{-\frac{1}{2}}}.\end{equation} 
At limit $q\to 1$, the {\it basic $q$-deformed quantum number} $[x]$ is reduced to the {\it number} $x$. 
In our present study we shall not attempt to explicitly use Jackson derivative $(JD)$ \cite{jak,jac}  to obtain $q$-deformed thermodynamics relations \cite{bri,lav1}. Instead, we start with the $q$-deformed partition function and use ordinary derivatives to obtain $q$-deformed thermodynamics quantities.
\section{Implementation of the $q$-deformation}
\label{iqd}
\subsection {$q$-Deformed Einstein solid}
\label{qde}
\begin{figure}
\includegraphics[{angle=90,height=8.0cm,angle=270,width=5.3cm}]{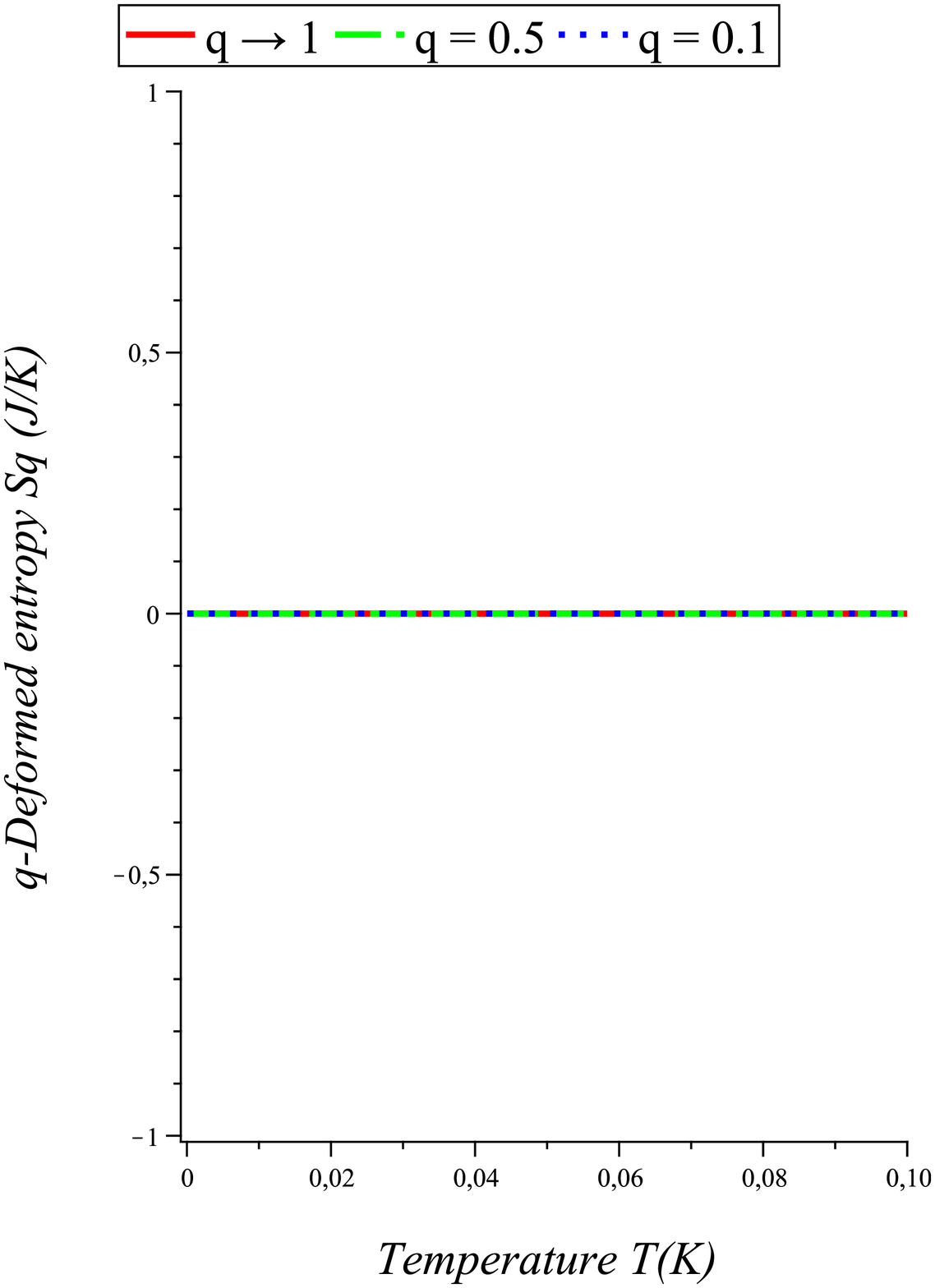} 
\includegraphics[{angle=90,height=8.0cm,angle=270,width=5.3cm}]{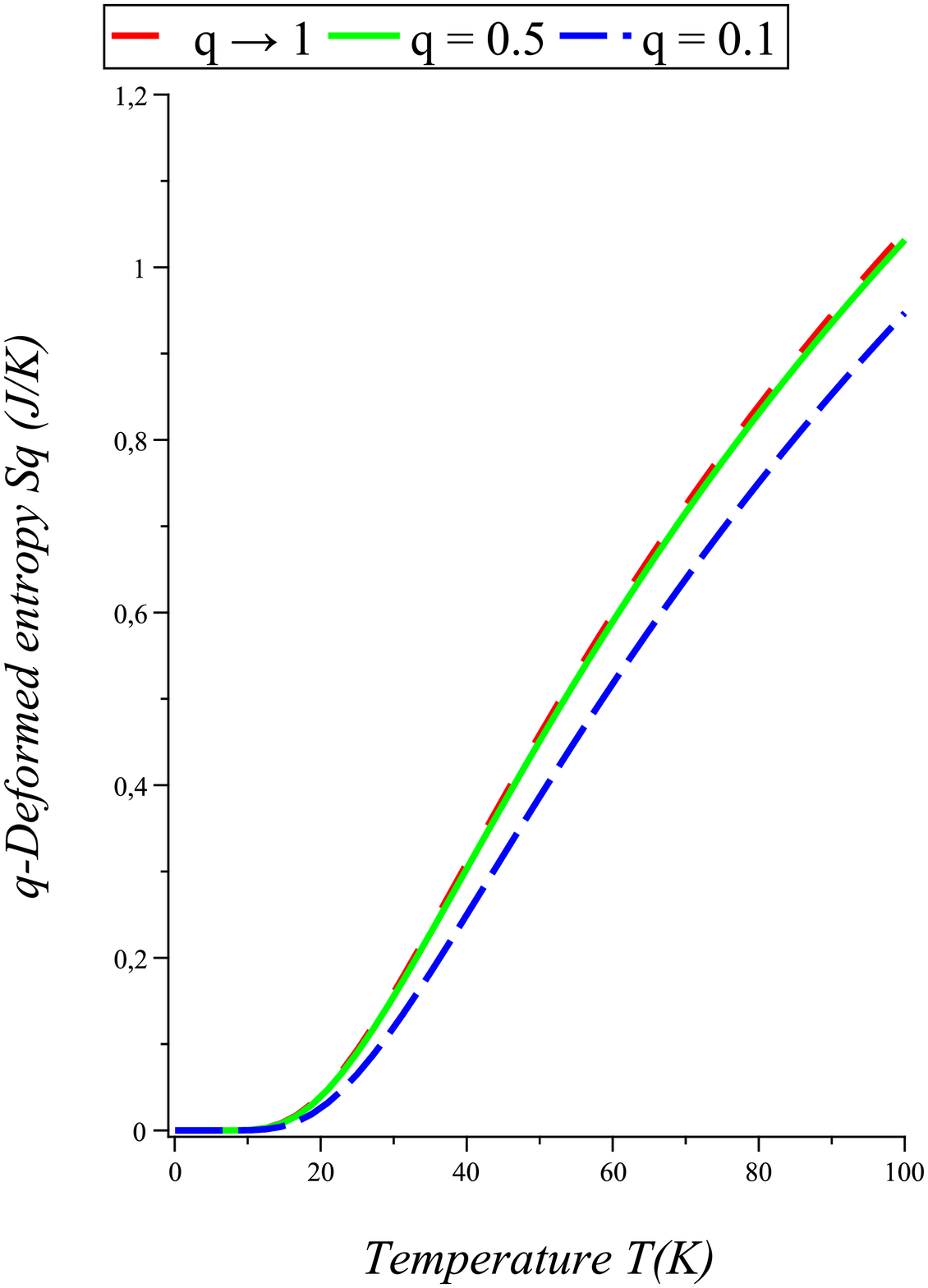}
\includegraphics[{angle=90,height=8.0cm,angle=270,width=5.3cm}]{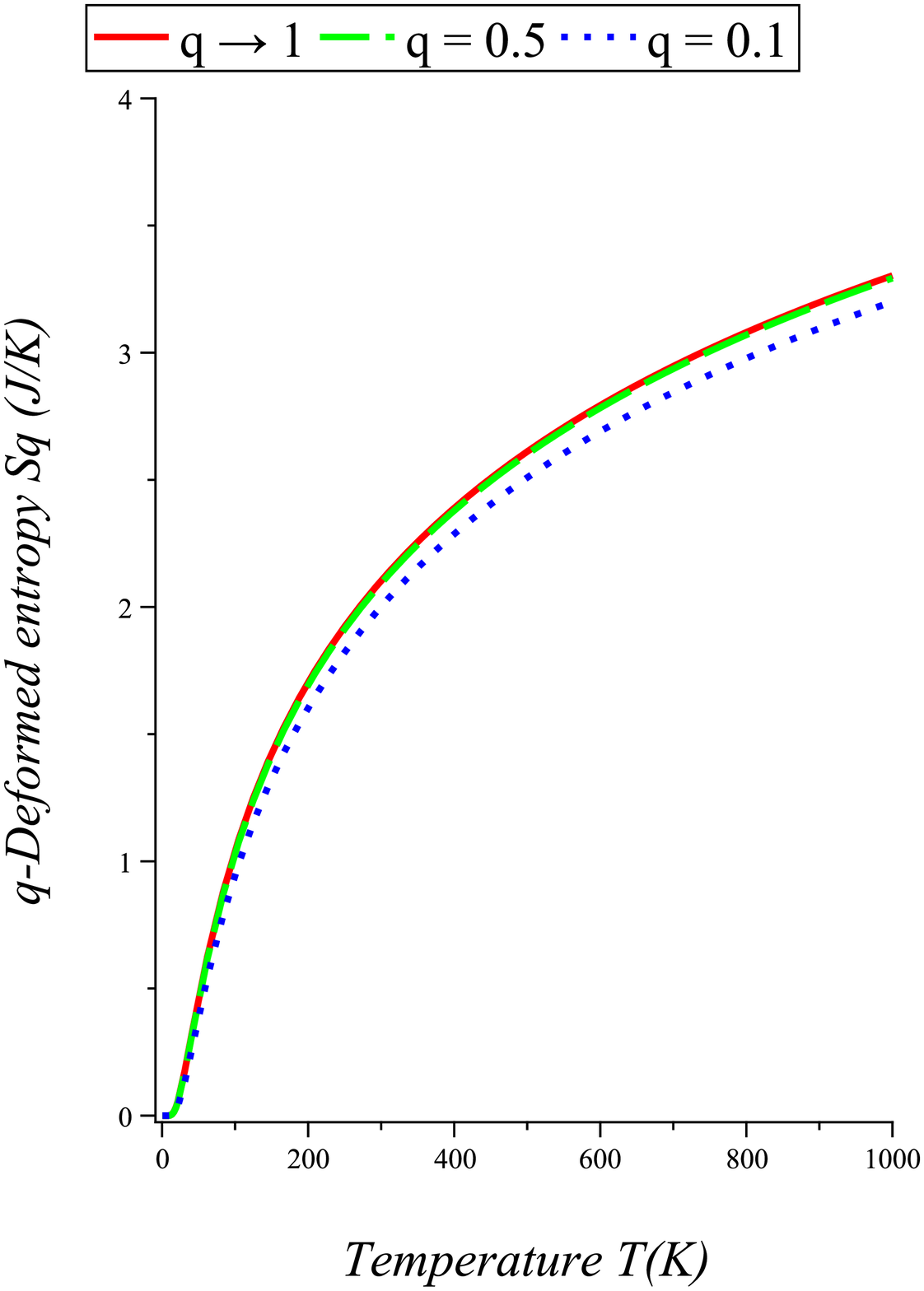} 
\caption{\footnotesize{$q$-Deformed entropy $S_q$ vs temperature $T$ in the following intervals: $T=0$ to $T=0.1 K $ \textbf{(top)}, 
$T=0$ to $T=100 K$ \textbf{(center)} and $T=0$ to $T=1000 K$ \textbf{(bottom)}}.}
\label{figura1}
\end{figure}

\begin{figure}
\includegraphics[{angle=90,height=8.0cm,angle=270,width=5.3cm}]{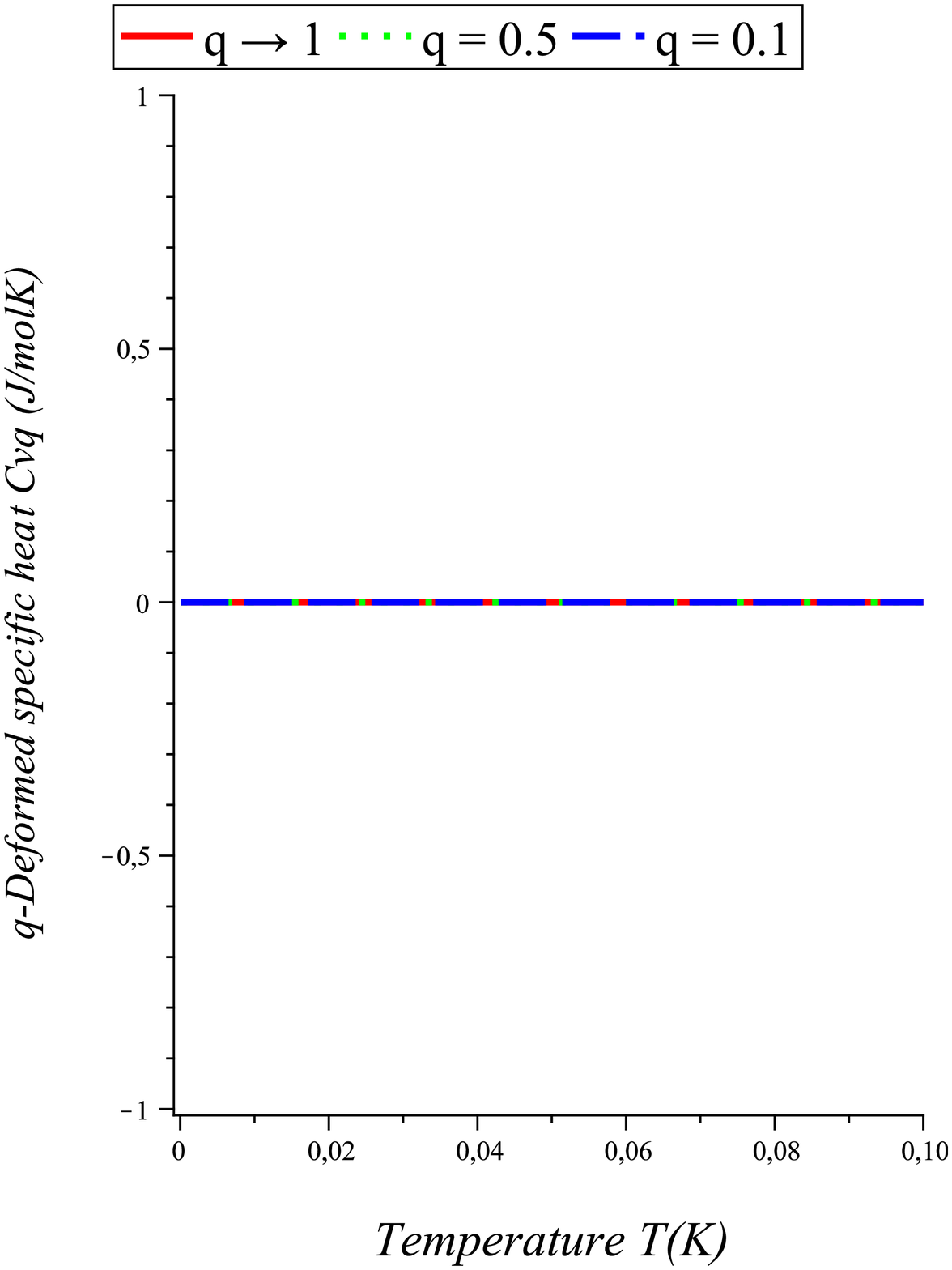} 
\includegraphics[{angle=90,height=8.0cm,angle=270,width=5.3cm}]{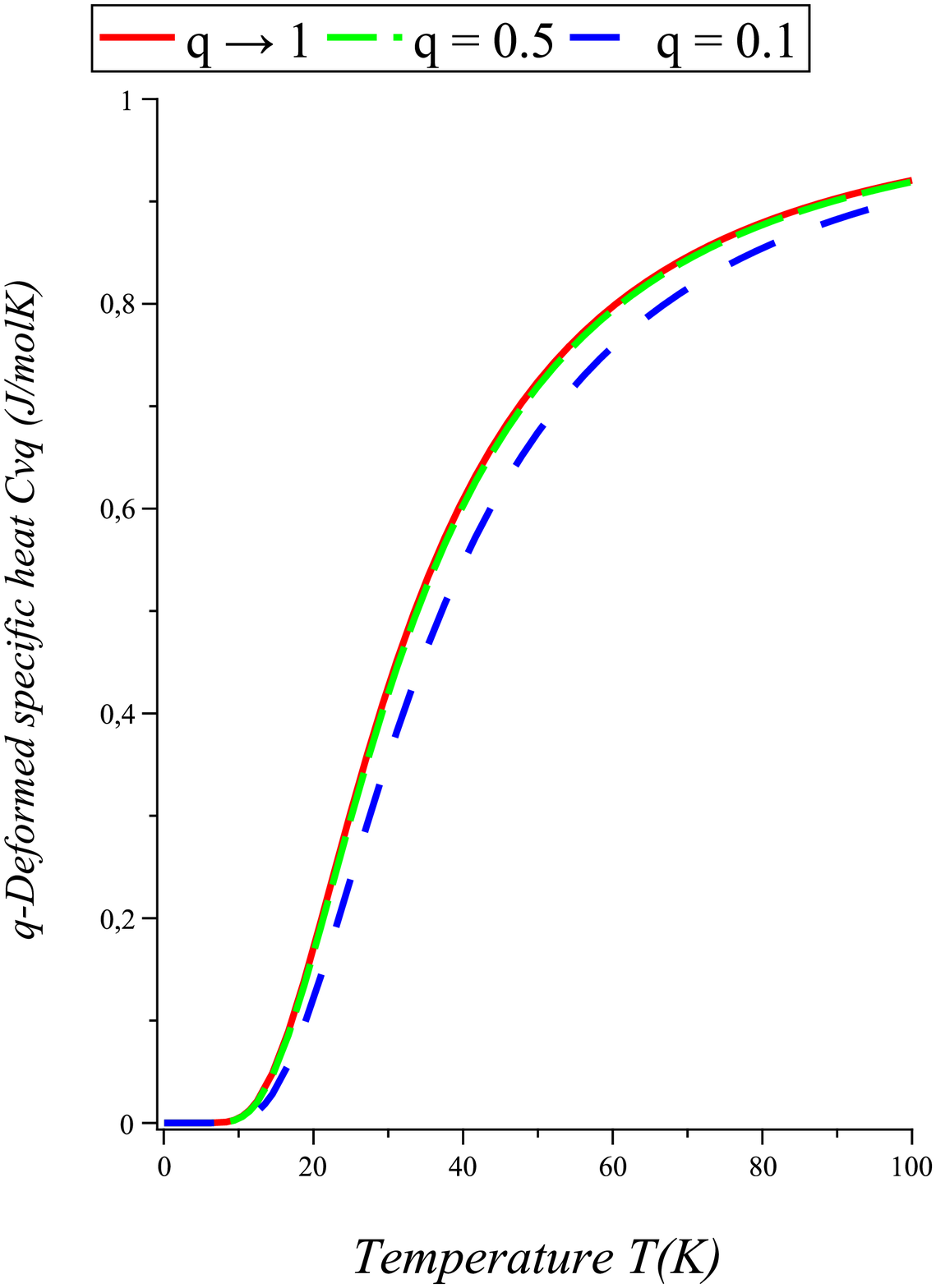}
\includegraphics[{angle=90,height=8.0cm,angle=270,width=5.3cm}]{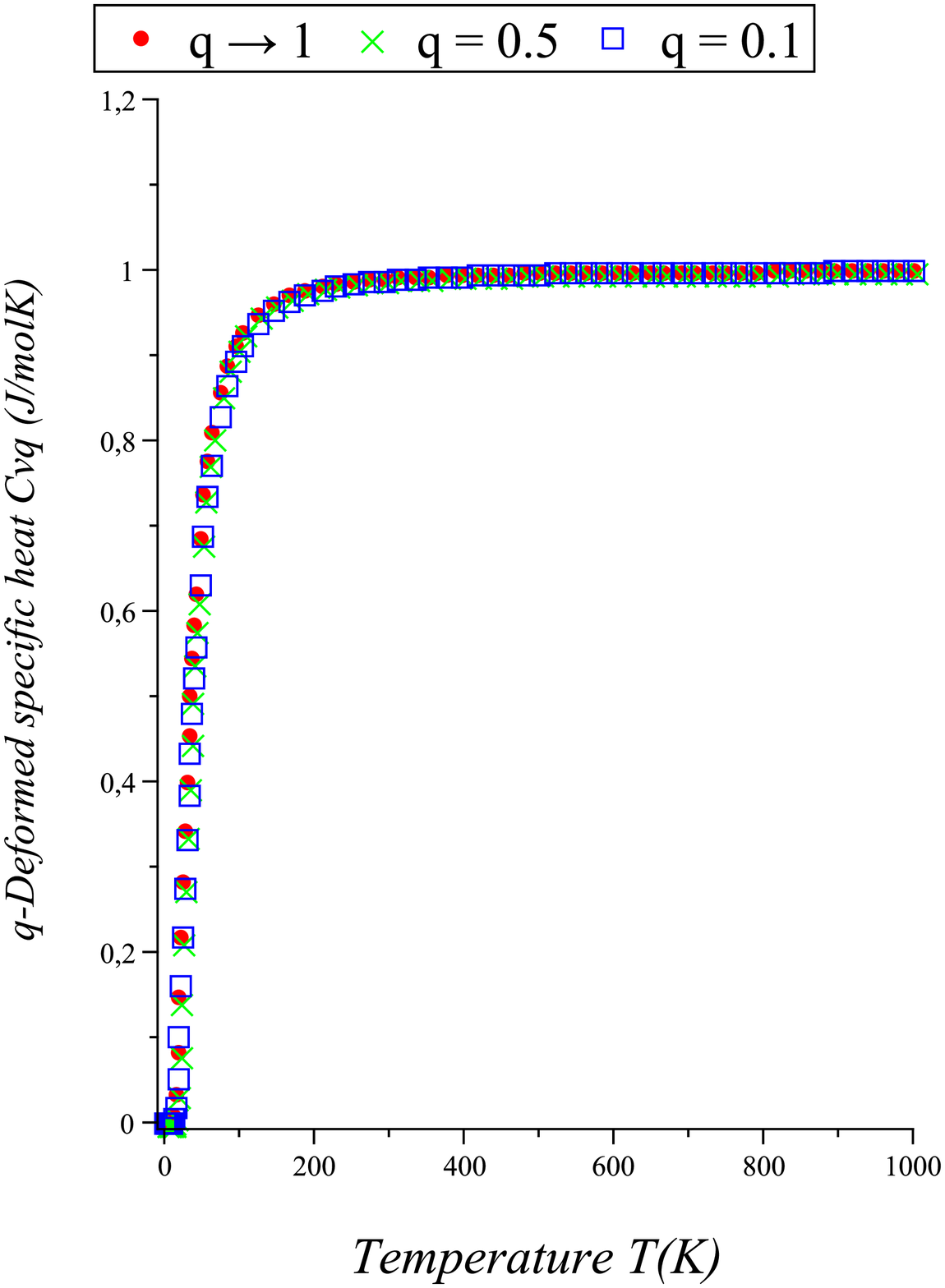} 
\caption{\footnotesize{$q$-Deformed specific heat $c_{Vq}$ vs temperature $T$ in the following intervals: $T=0$ to 
$T=0.1 K $ \textbf{(top)}, $T=0$ to $T=100 K$ \textbf{(center)} and $T=0$ to $T=1000 K$ \textbf{(bottom)}}.}
\label{figura11}
\end{figure}

We consider the solid in contact with a thermal reservoir at temperature $T$, where we have $n_j$ 
labeling the $j$-th oscillator. Given a microscopic 
state $\{n_j\}=\{n_1,n_2,\ldots,n_N\}$, the energy of this state can be written as 
\begin{equation}\label{eq8} E\{n_j\} = 
\displaystyle\sum_{j=1}^{\infty}\left(n_j+\frac{1}{2}\right)\hbar\omega_E,\end{equation}
Where $\omega_E$ is the Einstein frequency characteristic. As our primary purpose is 
$q$-deformation, we proceed to find the $q$-deformed version of (\ref{eq8}).
We can obtain $q$-deformed energies from the definition of the Hamiltonian (\ref{eq1}), and the definitions 
provided earlier, 
\begin{equation} \label{eq3}E_n = \frac{\hbar\omega_E}{2}\Big([n]+[n+1]\Big).\end{equation}
Considering the definition of {\it basic number} given in (\ref{eq2}), and making $q=\exp(\gamma)$, 
we obtain \cite{flo} 
\begin{equation} \label{eq4} [n] = 
\frac{\sinh\left({\frac{n\gamma}{2}}\right)}{\sinh\left(\frac{\gamma}{2}\right)}. \end{equation} 
Now replacing the equation (\ref{eq4}) into equation (\ref{eq3}), we have
\begin{equation} \label{eq5}E_{nq} = \frac{\hbar\omega_E}{2}\left[\frac{\sinh\left(\left(n+\frac{1}{2}\right)
\frac{\gamma}{2}\right)}{\sinh\left(\frac{\gamma}{4}\right)}\right]. \end{equation}
The partition function is given by 
\begin{equation} \Xi = \displaystyle\sum_{n_j} \exp(-\beta E\{n_j\}). \end{equation} 
Since there is no interaction terms, we factored the sum and apply the result of equation (\ref{eq5})
to obtain the $q$-deformed partition function 
\begin{equation} \Xi_{q} = \Bigg\{\displaystyle\sum_{n=0}^{\infty} \exp\left[-\beta E_{nq}
\right]\Bigg\}^N = \Xi_{1_q}^{N}, \end{equation} 
where
\begin{equation} \Xi_{1_q} = 
\frac{q^{\alpha_q}\exp\left(-2\alpha_q\left(\frac{\sqrt{q}-1}{\sqrt{q}+1}\right)\right)}
{q^{\alpha_q}-1}.\end{equation}
As one knows $\beta=\frac{1}{\kappa_{B}T}$ and $\Theta_E$ is the Einstein temperature, 
defined by 
\begin{equation} \Theta_E = \frac{\hbar\omega_E}{\kappa_B},\end{equation}
and we define a $q$-deformed Einstein function $E(\alpha_q)$ as 
\begin{equation}\label{eq9.1} E(\alpha_q) = q^{\alpha_q}\left(\frac{\alpha_{q}\gamma}
{q^{\alpha_q}-1}\right)^2, \end{equation}
\begin{equation} \alpha_q 
=\frac{\Theta_{E}(\sqrt{q}+1)}{8T\sqrt[4]{q}\sinh\left(\frac{\gamma}{4}\right)}\equiv \frac{\Theta_{E}}{4T}\frac{\exp{(\frac{\gamma}{2})}+1}{\exp{(\frac{\gamma}{2})}-1}.\end{equation}
We can determine a $q$-deformed Helmholtz free energy per oscillator from the expression
\begin{eqnarray}\label{eq10}
&\!\!\!\!\!\!\!\!f_q &\!=\! -\frac{1}{\beta}\displaystyle\lim_{N\to\infty}\frac{1}{N}\ln\Xi\nonumber\\
&\!\!\!\!\!\!\!\!=\!\!\!&\kappa_{B} 
T\Bigg\{2\alpha_q\!\left(\frac{\sqrt{q}-1}{\sqrt{q}+1}\right)-\ln(q^{\alpha_q})+\ln(q^{\alpha_q}-1\!)\!\Bigg\}.\end{eqnarray}
By using the result of equation (\ref{eq10}) we can determine the $q$-deformed entropy (see Fig.~\ref{figura1})
\begin{eqnarray}&S_q& = -\frac{\partial f}{\partial T}\nonumber\\
& = &
\kappa_{B}\Bigg\{\ln(q^{\alpha_q})-\ln(q^{\alpha_q}-1)+\frac{\alpha_q\ln(q)}{q^{\alpha_q}-1}\Bigg\}.\end{eqnarray} 
The $q$-deformed specific heat can now be determined by 
\begin{eqnarray}c_{Vq}(T) &=& T\left(\frac{\partial S_q}{\partial T}\right) \nonumber\\
& = &
\kappa_{B}q^{\alpha_q}\left(\frac{\ln(q)\alpha_q}{q^{\alpha_q}-1}\right)^2.\end{eqnarray} 
The specific heat of the Einstein solid as a function of $E(\alpha_q)$, defined in equation (\ref{eq9.1}), 
can be written as follows:

\begin{equation} \label{eq11} c_{Vq}(T) = 3\kappa_{B}E(\alpha_q).\end{equation} 
The complete behavior is depicted in Fig.~\ref{figura11}. One should note that when $T\gg\theta_E$, with the ratio 
$\alpha=\frac{\theta_E}{T}\ll 1$, 
with $\theta_E$, for example, around $100 K$ for common crystals, one recovers the classical result 
$c_{Vq}\to 3\kappa_B$, known as the Dulong-Petit law. However, for sufficiently low temperatures, 
where $T\ll\theta_E$ and therefore $\alpha\gg 1$, specific heat is exponential with 
temperature \cite{patt} as
\begin{equation}\label{eq15} c_{Vq} \rightarrow \kappa_B 
\left(\frac{\theta_E}{T}\right)^2\exp\left(-\frac{\theta_E}{T}\right). \end{equation} 
In general, the invariance of specific heat at high temperatures and its decrease at low temperatures 
show that the Einstein model is in agreement with experimental results. However, at sufficiently low 
temperatures, specific heat does not experimentally follow the exponential function given in equation 
(\ref{eq15}). 
As for the $q$-deformed case we see a significant change in the curve when $q\to 0.1$ for intermediate 
temperatures. However, we will improve the model by following the Debye model and applying $q$-deformation 
in the next section. 

Finally, the $q$-deformed internal energy per oscillator as a function of temperature is given through
\begin{eqnarray} u_{q} &=& -\frac{\partial}{\partial \beta}\ln\Xi_{1_q}\qquad\qquad\nonumber\\
 & =& \kappa_BT\left[2\alpha_q\left(\frac{\sqrt{q}-1}{\sqrt{q}+1}\right) + 
\frac{\alpha_q\ln(q)}{(q^{\alpha_q}-1)}\right].\end{eqnarray}
\subsection{$q$-Deformed Debye solid}
\label{qdd}
Corrections of Einstein's model are given by the Debye model, allowing us to integrate from 
a continuous spectrum of frequencies up to the Debye frequency $\omega_D$, giving the total number 
of normal modes of vibration \cite{patt,hua,kit,reif}
\begin{equation} \displaystyle\int_{0}^{\omega_D}g(\omega)d\omega = 3N, \end{equation}
where $g(\omega)d\omega$ denotes the number of normal modes of vibration whose frequency is in the range 
$(\omega, \omega+d\omega)$. The function $g(\omega)$, can be given in terms of Rayleigh expression as following 
\begin{equation} \label{eq.20}8\pi\left(\frac{1}{\lambda}\right)^{2}d\left(\frac{1}{\lambda}\right) = 
\frac{\omega^2 d\omega} {\pi^{2}c^{3}},\end{equation} 
where $c$ is the speed of light and $\lambda $ wavelength. The expected energy value of 
the Planck oscillator  with frequency $\omega_s$ is
\begin{equation}\label{eq.21}<E_s> = \frac{\hbar\omega_s}{\exp\left(\frac{\hbar\omega_s}{\kappa_{B}T}\right)-1}. 
\end{equation}

Using equations (\ref{eq.20}) and (\ref{eq.21}), we obtain the energy density associated to the frequency range
$(\omega,\omega+d\omega)$, 
\begin{equation} u(\omega)d\omega = \frac{\hbar}{\pi^2 c^3}\frac{\omega^3d\omega}{\exp\left(\frac{\hbar\omega} 
{\kappa_B T} \right)-1}.\end{equation}
To obtain the number of photons between $\omega$ and
$\omega+d\omega$, one makes use of the volume of the region on the phase space \cite{patt}, 
which results in 
\begin{eqnarray} \label{eq.22} g(\omega)d\omega \approx \frac{2V}{h^3}\left[4\pi\left(\frac{\hbar\omega}{c}\right)^2 
\left(\frac{\hbar d\omega}{c}\right)\right] = \frac{V\omega^{2}d\omega}{\pi^{2}c^{3}}. \end{eqnarray}

Thus, replacing the equation (\ref{eq.22}) into equation (\ref{eq.20}), we can write the specific heat 
for any temperature. 
We now apply $q$-deformation in the same way as in equation (\ref{eq11}), 
\begin{equation} c_{V_q}(T) = 3\kappa_{B}D(\alpha_0)_q, \end{equation}
where $D(\alpha_0)_q$ is the $q$-deformed Debye function, defined by 
\begin{equation} \label{eq.23} D(\alpha_0)_q = \frac{3}{\alpha_{0_q}^3} \int_{0}^{\alpha_{0_q}} 
\frac{\alpha^4\exp{\alpha}} 
{[\exp(\alpha)-1]^2}d\alpha, \end{equation} 
with
\begin{equation}\alpha_{0_q} = \frac{\hbar\omega_{D_{q}}}{\kappa_{B}T} = \frac{\theta_{D_{q}}}{T},\end{equation} 
and
\begin{equation} \omega_{D_{q}} = \frac{\omega_D\sinh\left(\frac{\ln(q)}{2}\right)} 
{2\sinh\left(\frac{\ln(q)}{4}\right)}, \end{equation} 
where $\omega_{D_{q}}$ is the $q$-deformed Debye frequency and $\theta_{D_{q}}$ is the $q$-deformed Debye temperature. 
Integrating equation (\ref{eq.23}) by parts one finds
\begin{eqnarray} D(\alpha_{0})_q = -\frac{3\alpha_{0_q}}{\exp(\alpha_{0_q})-1} + 
\frac{12}{\alpha_{0_q}^3}\int_{0}^{\alpha_{0_q}} 
{\frac{\alpha^3 d\alpha}{\exp(\alpha)-1}},\end{eqnarray} 
that can be integrated out to give the full expression
\begin{eqnarray}\label{eq-cVq} &&\!\!\!\!\!D(\alpha_{0})_q=\nonumber\\
&\!\!\!\!\!=\!\!\!\!\!& 3\kappa_{B}\Biggl\{-\frac{3\alpha_{0_q}}{\exp{\left(\alpha_{0_q}\right)}-1}+\frac{12}{\alpha_{0_q}^{3}}
\Bigg[-\frac{\pi^{4}}{15}-\frac{1}{4}\alpha_{0_q}^{4}+\nonumber\\
&\!\!\!\!\!+\!\!\!\!\!&\alpha_{0_q}^{3}\ln\left[1-\exp\left(\alpha_{0_q}\right)\right] 
+3\left(\alpha_{0_q}\right)^{2}{\rm Li}_2\left[\exp\left(\alpha_{0_q}\right)\right] -\nonumber\\
&\!\!\!\!\!-\!\!\!\!\!&6\left(\alpha_{0_q}\right){\rm Li}_3\left[\exp\left(\alpha_{0_q}\right]\right)
+6\ {\rm Li}_4\left[\exp\left(\alpha_{0_q}\right)\right] 
 \Bigg]\Biggl\},\end{eqnarray}
where 
\begin{eqnarray}
{\rm Li}_n(z)={\sum_{k=1}^\infty}\frac{z^k}{k^n}
\end{eqnarray}
is the polylogarithm function.
\begin{figure}[hb!]
\includegraphics[{angle=90,height=8.0cm,angle=270,width=8.0cm}]{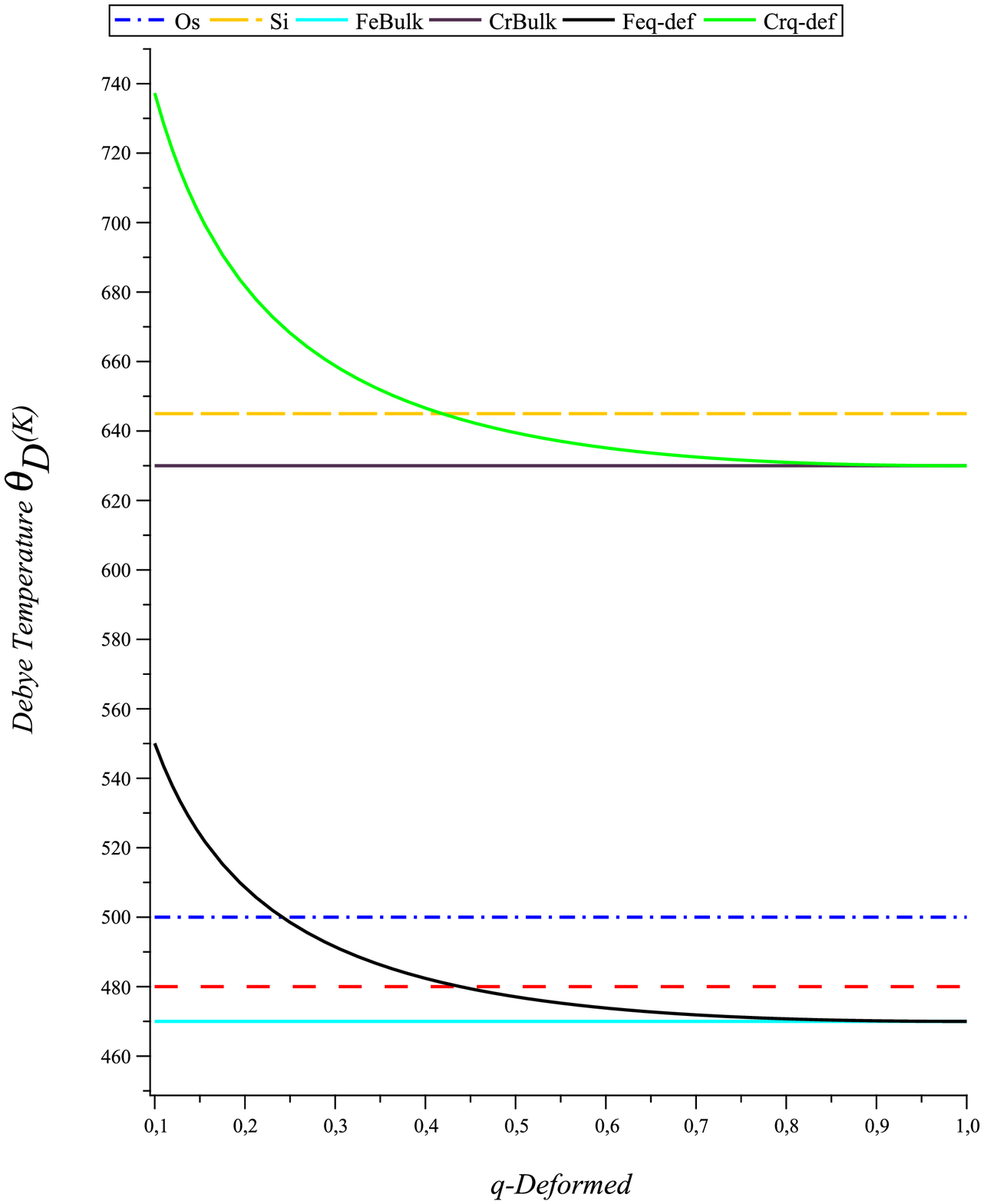} 
\includegraphics[{angle=90,height=8.0cm,angle=270,width=8.0cm}]{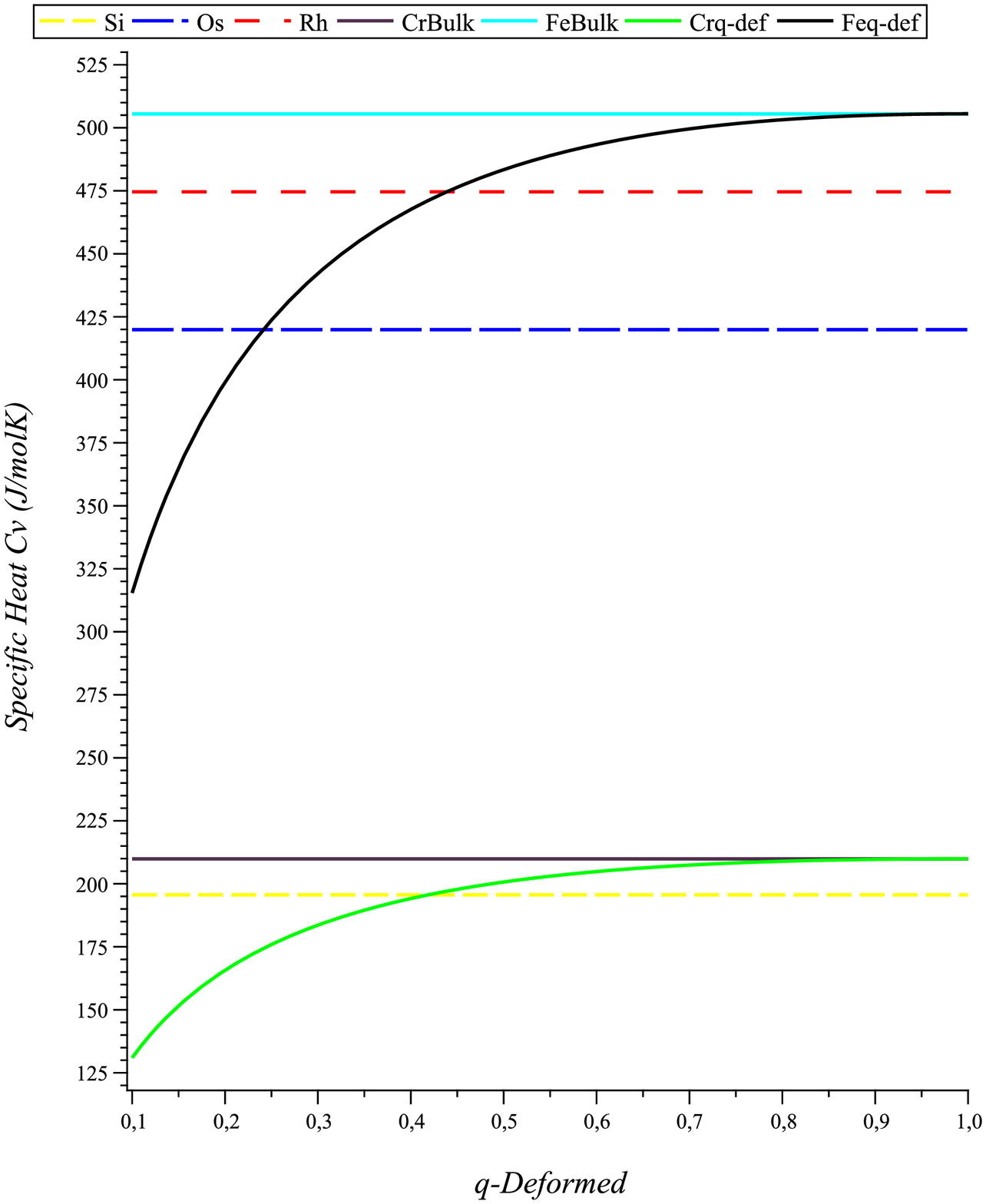}
\caption{\scriptsize{Debye temperature $\theta_D$ range of $Fe$ and $Cr$ varying as a function 
of $q=0.1$ to $q\to 1$ \textbf{(top)}; $q$-deformed specific heat $c_V$ range of $Fe$ and $Cr$ varying 
as a function of $q=0.1$ to $q\to 1$ \textbf{(bottom)}}.}
\label{figura2}
\end{figure}
For $T\gg\theta_{D_{q}}$, then $\alpha_{0_q}\ll 1$, the function $D(\alpha_0)_q$ can be expressed in a power series in 
$\alpha_{0_q}$ 
\begin{equation} D(\alpha_{0})_q = 1 - \frac{\alpha_{0_q}^2}{20} + \cdots \end{equation}
so that for
\begin{eqnarray} T\to\infty\;,\qquad c_{V_q}\to 3\kappa_B.
\end{eqnarray}
For $T\ll \theta_{D_{q}}$, then $\alpha_{0_q}\gg 1$, we can write function $D(\alpha_0)_q$ as 
\begin{equation} 
\frac{12}{\alpha_{0_q}^{3}}\displaystyle\int_{0}^{\infty}{\frac{\alpha^3d\alpha}{\exp(\alpha)-1}+O[\exp(-\alpha_{0_q})
]}, 
\end{equation}
\begin{equation}\approx \frac{4\pi^4}{5\alpha_{0_q}^{3}} = \frac{4\pi^4}{5}\left(\frac{T}{\theta_{D_{q}}}\right)^3. 
\end{equation}

Thus, as in the usual Debye solid, low temperature specific heat in a {\it $q$-deformed Debye solid} is proportional to $T^3$ rather than
proportional to the exponential function in (\ref{eq15}) for the $q$-deformed Einstein solid. 
This is in agreement with experiments. Finally, we express the specific heat for low temperatures as 
\begin{eqnarray} c_{V_q} = \frac{12\pi^4\kappa_B}{5}\left(\frac{T}{\theta_{D_{q}}}\right)^3 = 1944\left(\frac{T} 
{\theta_{D_{q}}}\right)^3 \frac{J}{mol K}.
 \end{eqnarray}
For the $q$-deformed case one can observe the changes that occur with Debye temperature, specific heat, thermal and
electrical conductivies and electric resistivity. 
By using the relationship established for thermal 
conductivity $\kappa$ \cite {zim} we obtain 
\begin{equation} \label{eq23}\kappa = \frac{1}{3}C_{V}v l, \end{equation} 
where $v$ is the average velocity of the particle, $C_{V}$ is the molar heat capacity and $l$ is the space 
between particles. We can deduce a relationship between the thermal and electrical $\sigma$ conductivities through the 
elimination of $l$ (as $\sigma=\frac{ne^2l}{mv}$, where $m$ is the electron mass, $n$ is the number of electrons per volume 
unit and $e$ is the electron charge), such that 
\begin{equation} \frac{\kappa}{\sigma}=\frac{1}{3}\frac{C_{V}mv^2}{ne^2}.\end{equation}
\begin{figure}[hb!]
\centerline{
\includegraphics[{angle=90,height=8.0cm,angle=270,width=8.0cm}]{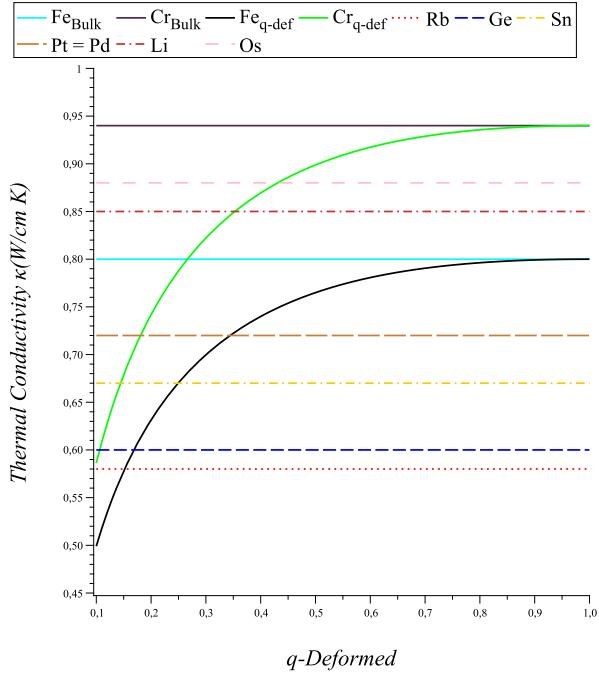} 
}\caption{\footnotesize{Thermal conductivity $\kappa$ range of $Fe$ and $Cr$ varying as a function 
of $q=0.1$ to $q\to 1$.}}
\label{figura3}
\end{figure}
In a classical gas the average energy of a particle is $\frac{1}{2}mv^2=\frac{3}{2}\kappa_{B}T$, 
whereas the heat capacity is $\frac{3}{2}n\kappa_{B}$, so that 
\begin{equation}\label{eq24} \frac{\kappa}{\sigma}=\frac{\pi^{2}}{3}\left(\frac{\kappa_B}{e}\right)^2T 
\end{equation}
the ratio $\frac{\kappa}{\sigma T}$ is called the \textit{Lorenz number} and should be a constant, 
independent of the temperature and of the scattering mechanism. This is the famous Wiedemann-Franz law, 
which is often well satisfied experimentally, and the Lorenz number correctly given \cite{zim}. 
\begin{figure}[hb!]
\includegraphics[{angle=90,height=8.0cm,angle=270,width=8.0cm}]{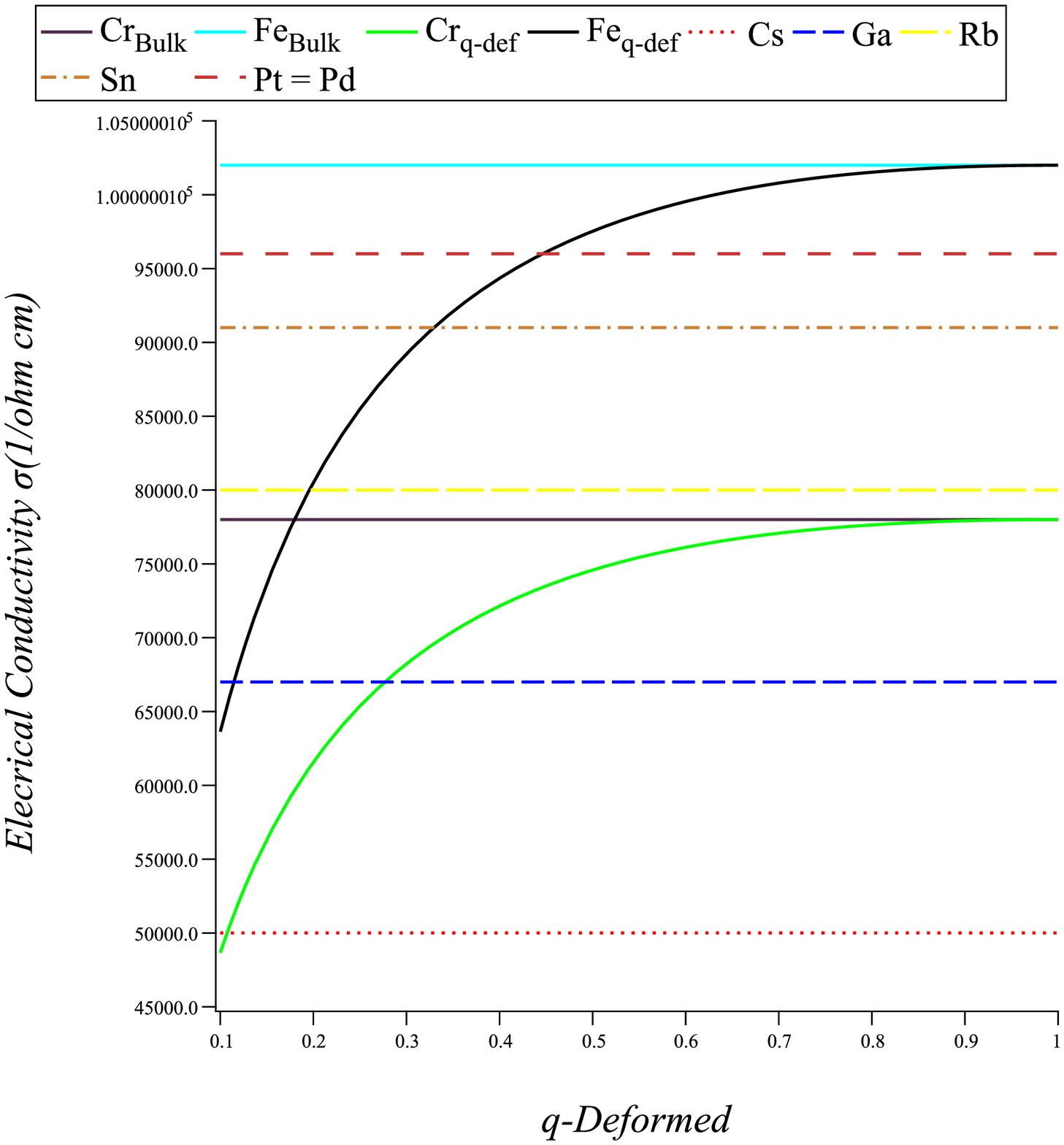} 
\includegraphics[{angle=90,height=8.0cm,angle=270,width=8.0cm}]{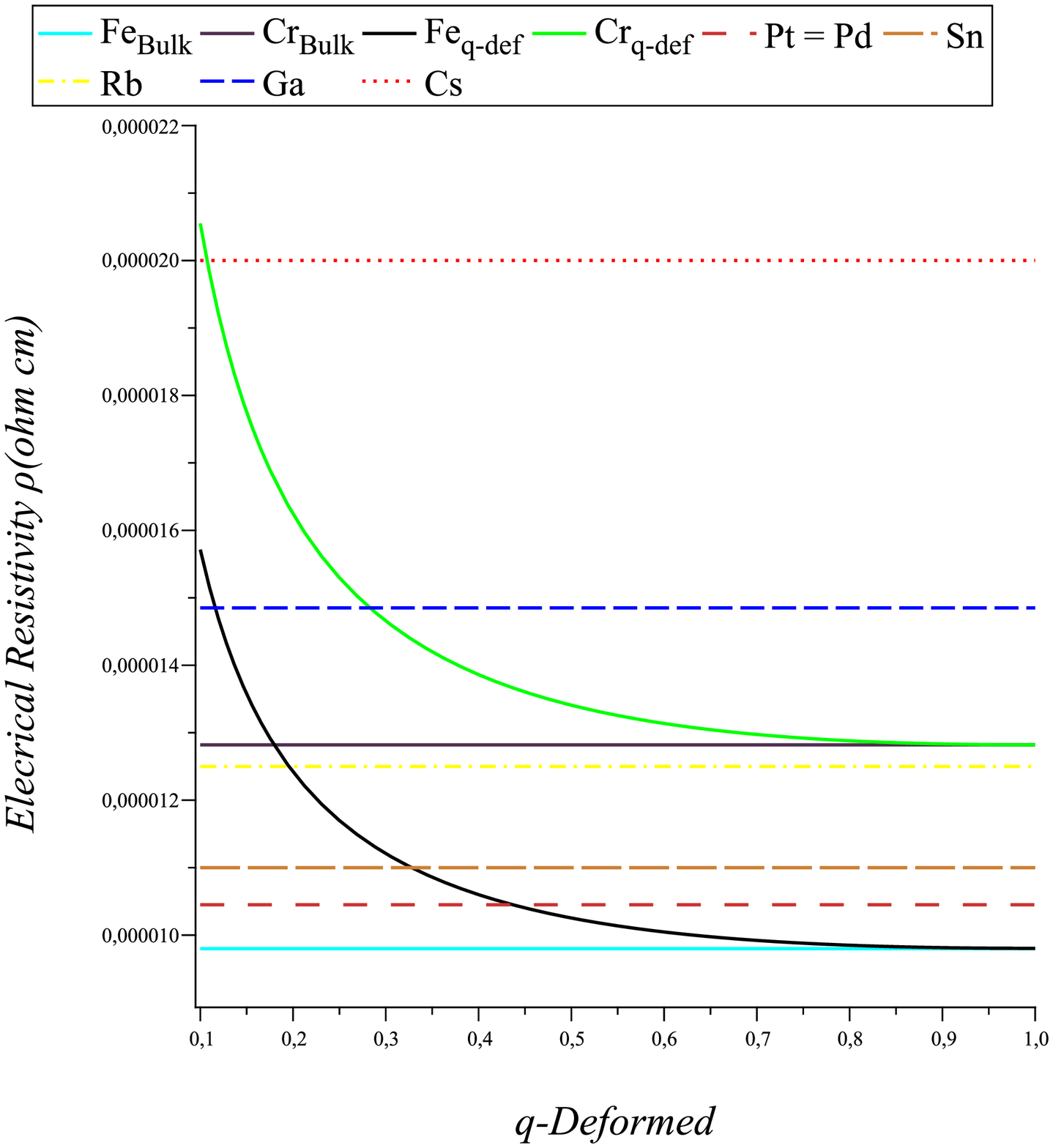}
\caption{\footnotesize{Electrical conductivity $\sigma$ range of $Fe$ and $Cr$ varying as a function 
of $q=0.1$ to $q\to 1$ \textbf{(top)}; electrical resistivity $\rho$ range of $Fe$ and $Cr$ varying as a function
of $q=0.1$ to $q\to 1$ \textbf{(bottom).}}}
\label{figura4}
\end{figure}
By using the $q$-deformed relations presented above, we start from Eqs.~(\ref{eq23}) and (\ref{eq24}) to determine 
the important relations for $q$-deformed thermal and electrical conductivities 
\begin{eqnarray}\kappa_q=\frac{\kappa c_{V_q}}{c_{V}} \qquad\qquad \mbox{and} \qquad\qquad 
\sigma_q=\frac{\kappa_{q}\sigma}{\kappa}.\end{eqnarray}
Since electrical resistivity $\rho$ is the inverse of conductivity, $q$-deformed 
$\rho_{q}=\frac{1}{\sigma_{q}}$. Recall that to compute these deformed quantities in terms of the 
specific heat $c_{Vq}$ we make use of the equation (\ref{eq-cVq}) and its suitable limits. 
Data is plotted to provide a better view of our results. We see how $q$-deformation 
is acting on chemical element properties. For illustration purposes, we chose iron (Fe) and chromium 
(Cr) two materials that can be employed in many areas of interest.

Fig.~\ref{figura2} shows how $q$-deformation acts on $Fe$ and $Cr$ Debye temperatures and specific heat.
The plots show that $Fe$ reaches  rhodium (Rh) values for $q\approx 0.45$, osmium 
(Os) for $q\approx 0.25$ while $Cr$ approaches silicon (Si) for $q\approx 0.42$. 

Fig.~\ref{figura3} shows that thermal conductivity for $Fe$ approaches the thermal conductivity of rubidium (Rb) for 
$q\approx 0.15$, whereas $Cr$ equals bulk $Fe$ for $q\approx 0.27$, lithium (Li) for 
$q\approx 0.35$ and $Os$ for $q\approx 0.48$. They also approach the germanium (Ge) for $q\approx 0.17$ 
and $q\approx 0.1$, tin (Sn) for $q\approx 0.25$ and $q\approx 0.15$, platinum (Pt) and 
palladium (Pd) for $q\approx 0.35$ and $q\approx 0.18$, respectively.

Fig.~\ref{figura4} shows that electrical conductivity and resistivity of 
$Fe$ equals $Ga$ for $q\approx 0.12$, bulk $Cr$ for $q\approx 0.18$, $Rb$ for $q\approx 0.2$, 
$Sn$ for $q\approx 0.33$, $Pt$ and $Pd$ for $q\approx 0.45$, while $Cr$ equals cesium (Cs) with 
$q\approx 0.11$ and $Ga$ for $q\approx 0.29$. 
\begin{figure}
\includegraphics[{angle=90,height=8.0cm,angle=270,width=8.0cm}]{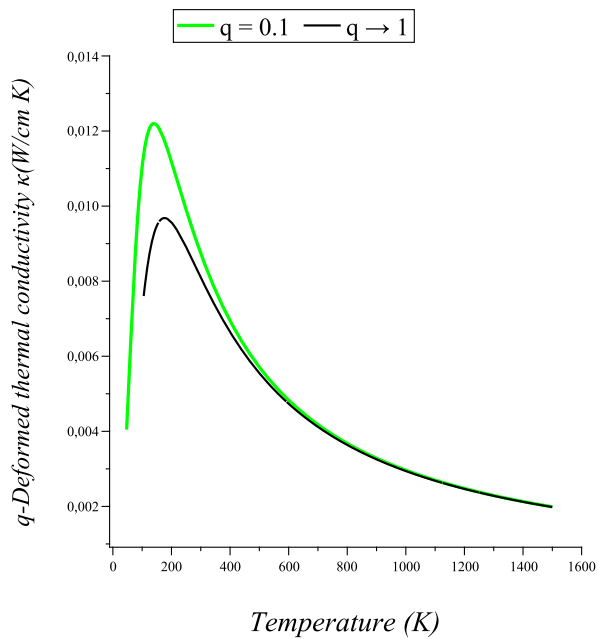} 
\caption{$q$-Deformed thermal conductivity $\kappa_q$ for $Cr$ as $q\to 1$ and for $Cr$ with impurities as $q=0.1$ vs temperature $T$, in the interval
 $T=0$ to $T=1500 K$ .}
\label{figura5}
\end{figure}

Fig.~\ref{figura5} shows correct behavior of thermal conductivity for a pure and impure material. One should note that $q$-deformation is clearly  playing the role of impurity concentration in the material sample. This is because $q$-deformation acts directly on Debye temperature, which means that the Debye frequency is modified. Changing the Debye frequency is a clear sign of the material being modified by impurities. 

\section{Conclusions}
\label{con}

Following our previous results in \cite{bri}, we understand $q$-deformation not only as a mathematical tool, but 
also as an impurity factor in a material, such as disorder or reorganization of a crystalline structure. 

In the present study, we first investigate the Einstein solid model to obtain thermodynamic quantities such as Einstein temperature, Helmholtz free energy, specific heat and internal energy with $q\neq 1$. We then generalize this study to the Debye solid model to obtain  the Debye temperature, specific heat, thermal conductivity, electrical conductivity and resistivity. We present these findings for a number of chemical elements.

Our main results indicate the possibility of adjusting $q$-deformation to obtain 
desirable physical effects, such as changing the
thermal conductivity of a certain element, which might become equivalent to a material that is easier to handle by inserting an impurity in a sample of the original element. We need further studies and 
evidence to substantiate such a complex hypothesis. For example, we are seeking to establish a connection between this theory and experiments through the growth of thin films, a matter that will be addressed elsewhere. 

\acknowledgments

We would like to thank CAPES, CNPq and PNPD/PROCAD-CAPES, for financial support.


\end{document}